\documentclass[a4paper,11pt]{article}
\usepackage{amsmath, amssymb, amsfonts,indentfirst,graphicx,color,anysize}
\marginsize{3cm}{3cm}{2cm}{2cm}
\usepackage{xspace}

\newcommand*{\Nch}{\ensuremath{N_\mathrm{ch}}\xspace}
\newcommand*{\Nfw}{\ensuremath{N_\mathrm{fw}}\xspace}
\newcommand*{\Nmpi}{\ensuremath{N_\mathrm{MPI}}\xspace}
\newcommand*{\Ntrans}{\ensuremath{N_\mathrm{trans}}\xspace}
\newcommand*{\Ncone}{\ensuremath{N_\mathrm{cone}}\xspace}
\newcommand*{\RT}{\ensuremath{R_\mathrm{T}}\xspace}
\newcommand*{\RNC}{\ensuremath{R_\mathrm{NC}}\xspace}
\newcommand*{\SO}{\ensuremath{S_\mathrm{0}}\xspace}
\newcommand*{\pT}{\ensuremath{p_\mathrm{T}}\xspace}

\newcommand*{\GeVc}{\ensuremath{\mathrm{GeV}/c}\xspace}

\newcommand*{\Lc}{\ensuremath{\Lambda_{\rm c}^+}\xspace}
\newcommand*{\Xic}{\ensuremath{\Xi_{\rm c}^{0,+}}\xspace}
\newcommand*{\Sigc}{\ensuremath{\Sigma_{\rm c}^{0,++}}\xspace}
\newcommand*{\Omc}{\ensuremath{\Omega_{\rm c}^{0}}\xspace}
\newcommand*{\Dz}{\ensuremath{\rm D^0}\xspace}

\newcommand*{\LcToDz}{\ensuremath{{\Lambda_{\rm c}^+}/{\rm D^0}}\xspace}
\newcommand*{\SigcToDz}{\ensuremath{{\Sigma_{\rm c}^{0,++}}/{\rm D^0}}\xspace}

\title{
	\includegraphics[width=0.35\textwidth]{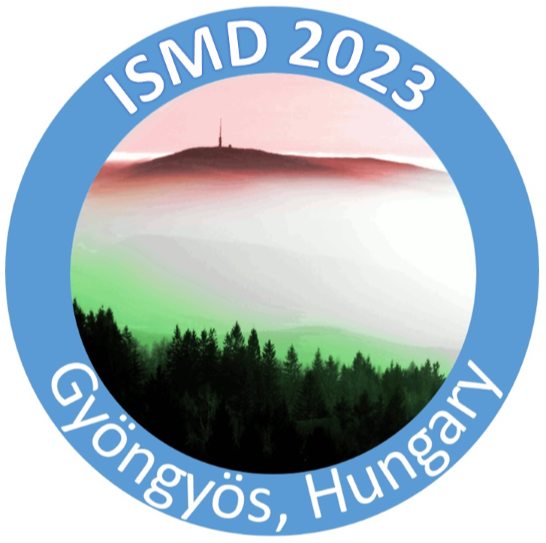}\\[1cm]
	\textbf{Connection of event shapes to the heavy-flavor baryon enhancement}}
\author{{R. Vértesi$^{1}$ and Z. Varga$^{1,2}$,}\\[1ex]
	$^1$HUN-REN Wigner Research Centre for Physics,\\
	29-33 Konkoly-Thege Miklós út, 1121 Budapest, Hungary\\
	$^2$Budapest University of Technology and Economics,\\
	3 Műegyetem Rakpart, 1111 Budapest, Hungary\\
}
\begin{document}

\maketitle

\begin{abstract} 
Recent results from ALICE and CMS show a low-transverse-momentum enhancement of charm baryon-to-meson production ratios over model predictions based on e$^+$e$^-$ collisions. This new development challenges the universality of fragmentation. We studied the charm-baryon enhancement in collision events generated by PYTHIA~8 and applied a color-reconnection model beyond leading color approximation. We proposed a measurement method based on several event-activity classifiers, to identify the origin of the charm-baryon enhancement. In this work we extend our studies to a new event classifier, flattenicity, that considers a broad pseudorapidity range. We have also studied the role of isospin and strangeness by comparing the production of different charmed baryons. The observables we explored provide a unique opportunity in the upcoming measurements from the high-luminosity LHC Run 3 period to better understand heavy-flavor fragmentation mechanisms, and will help the further development of models.
\end{abstract}

\section{Introduction}

Collective behavior observed in ultra-relativistic collisions of heavy ions has long been considered as a tell-tale sign of quark-gluon plasma. Surprisingly, small collisional systems, such as proton--proton (pp) or proton--lead (p--Pb), were found at the LHC to exhibit long-range multiparticle correlations.
An intensively researched question in high-energy physics is whether small droplets of quark-gluon plasma may come into being in collisions of small systems. Although the picture is still not entirely clear, a consensus tends to form that 
the observed collective phenomena can be explained by semi-soft vacuum-QCD effects, such as multiparton-interactions~\cite{Schlichting:2016xmj} with color-reconnection~\cite{Ortiz:2016kpz} or minijets (semi-hard partons produced by incoming partons or bremsstrahlung)~\cite{Eskola:1997au} production.

Charm baryon-to-meson ratios are sensitive probes of heavy-quark fragmentation. Heavy-flavor production is usually described using the factorization hypothesis, which assumes that the cross section of the heavy-flavor hadron production can be written as
\begin{equation}
	\sigma_{hh\rightarrow H} = f_{a}(x_1,Q^2) \otimes f_{b}(x_2,Q^2)
	 \otimes \sigma_{ab\rightarrow H} \otimes D_{q \rightarrow H}(z_q,Q^2) \,
\end{equation}
where $f_{a}$ and $f_{b}$ are the parton distribution functions of the incoming partons corresponding to a given Feynman-$x$ and momentum transfer $Q^2$, $\sigma_{ab\rightarrow H}$ is the cross-section of the hard scattering process, and $D_{q \rightarrow H}$ is fragmentation function corresponding to the heavy-flavor hadron $H$ depending on the quark-momentum fraction $z_q$. Fragmentation functions are traditionally treated as universal across all colliding systems. However, the ALICE experiment has recently found that the charmed baryon-to-meson ratios
\LcToDz and \SigcToDz are significantly underestimated by models based on factorization approach with fragmentation functions from
e$^+$e$^-$ collisions~\cite{ALICE:2021rzj}, suggesting that the heavy-flavor fragmentation universality is broken.
There are three main scenarios that intend to describe this discrepancy. 
String formation beyond leading color approximation~\cite{Christiansen:2015yqa} implemented in 
PYTHIA 8~\cite{Sjostrand:2014zea} (CR-BLC), statistical hadronization with feed-down from an augmented set of charmed-baryon states~\cite{He:2019tik}, and charm--light quark coalescence based models such as the Catania model with fragmentation and coalescence~\cite{Plumari:2017ntm} as well as a coalescence model based on statistical weights with equal quark-velocity~\cite{Song:2018tpv} all describe the trends observed in the momentum-dependence of \LcToDz ratios. More recently, ALICE observed that in the mid-tansverse-momentum regime $2<\pT<8$ GeV/$c$, the enhancement in the \LcToDz ratio increases with event multiplicity~\cite{ALICE:2021npz}.
It is to be noted that the \pT-integrated yields do not show this dependence on the activity of the event.

While \Lc is the most abundant among the charm baryons produced in high-energy collisions, other states can provide us with additional insight to production mechanisms. The \Sigc baryon has a quark content identical to that of \Lc (qqc), as these two only differ in their isospin. Observation of the  \Xic (qsc) and \Omc (ssc) baryons with both charm and strangeness content can reveal the possible connection between the excess production of charmed baryons and the strangeness enhancement observed in high-multiplicity events~\cite{ALICE:2016fzo}. Recent experimental results on the production of \Sigc, \Xic and \Omc baryons~\cite{ALICE:2021rzj,ALICE:2021bli,ALICE:2022cop,ALICE:2021psx} are only partially understood by the aforementioned models.

In the current work we summarize and extend our efforts aimed at understanding the origin of charm-baryon enhancement. We present detailed event activity dependent studies that examine whether the excess production is linked to the jet or the underlying event (UE). The first part of the paper we use event-activity and event-shape dependent observables and we test their sensitivity by examining the \LcToDz ratio in the CR-BLC scenario. These results extend our earlier work described in Ref.~\cite{Varga:2021jzb}. In the second part, which is based on Ref.~\cite{Varga:2023byp}, we examine the production of \Sigc, \Xic and \Omc within the CR-BLC model in order to understand the effect of isospin as well as strangeness content on the charmed-baryon enhancement.

\section{Simulation and analysis} 

As described in Ref.~\cite{Varga:2021jzb} more in details, we used PYTHIA 8.303~\cite{Sjostrand:2014zea} with soft-QCD settings and color reconnection beyond leading color approximation~\cite{Christiansen:2015yqa} to simulate 1 billion pp collision events at  $\sqrt{s}=13$ TeV.
The base of the CR-LBC modes is the Monash tune~\cite{Skands:2014pea}, a set of PYTHIA parameters that has been optimized to describe a broad set of minimum-bias, Drell--Yan and underlying-event data from the LHC to constrain the initial-state-radiation and multi-parton-interaction parameters, combined with data from SPS and the Tevatron LHC to constrain the energy scaling.
The starting point of our study was CR-LBC mode 2, which includes time dilation using the boost factor obtained from the final-state mass of the dipoles and requires a causal connection of all dipoles, and it is known to reproduce the trends in \LcToDz ratios well~\cite{ALICE:2020wfu,Hills:2021eto}.
As a cross-check we used mode 0, which uses no time-dilation constraints and where the amount of CR is controlled by the invariant mass scale parameter. The results yielded by these settings are qualitatively similar to those from mode 2~\cite{Varga:2021jzb}. We also investigated mode 3, which has time dilation constraints but only requires a single casual connection, and found that it significantly overestimates the underlying event and therefore we did not use it.

In our studies we roughly reproduced the experimental setup of the ALICE experiment. We selected final-state charged particles in the central pseudorapidity window $|\eta|<1$, in the full azimuth ($\varphi$) range. We required a minimum transverse momentum \pT$>0.15$~\GeVc to emulate experimental constraints. We used the Monte-Carlo particle identification codes~\cite{Zyla:2020zbs} to select 
\Lc, \Sigc, \Xic, $\Omega^{0}_{c}$ as well as the charmed \Dz meson and their charged conjugates within the rapidity window $|y|<0.5$. 

\section{Event classification}

In order to understand the mechanism of the enhancement in the charmed baryon-to-meson ratios, we investigated the \LcToDz ratio in terms of several event classifiers. 
The simplest event-activity classifiers are the central and forward charged-hadron multiplicities, \Nch and \Nfw, which we gained by counting the charged final-state hadrons in the central and forward pseudorapidity $|\eta|<1$ and $2<|\eta|<5$ respectively, roughly corresponding to the acceptance of ALICE. The charged-hadron multiplicities represent an event globally and are not selective for the jet or UE region.
To achieve such sensitivity, we applied the relative transverse multiplicity quantifier \RT~\cite{Martin:2016igp}, defined as 
\begin{equation}
	\RT = \frac{\Ntrans}{\left< \Ntrans \right>} \ , 
\end{equation}
where \Ntrans is the transverse charged-hadron multiplicity in an event, defined as the number of charged hadrons above $\pT=150$ MeV/$c$ in the range $\pi/2 < \Delta \varphi < 3\pi /2$, the angle difference defined in the transverse plane with respect to the leading (highest-\pT) charged hadron. The leading hadron is required to have a $\pT>5$ GeV/$c$, therefore \RT is only defined in a fraction of events. 
The value of \RT represents the multiplicity in the underlying event, and it is in close connection to \Nmpi in models like PYTHIA 8~\cite{Martin:2016igp}.

To describe the event activity in an analogous way to \RT, one can define the relative near-side cone multiplicity quantifier \RNC~\cite{Varga:2021jzb} as 
\begin{equation}
	\RNC = \frac{\Ncone}{\left< \Ncone \right>} \ , 
\end{equation}
where \Ncone is the charged-hadron multiplicity within a cone with a radius of $r = \sqrt{(\Delta \varphi^2 + \Delta \eta^2)} $, $\Delta\varphi$ and $\Delta\eta$ being the relative azimuth angle and pseudorapidity compared to the leading hadron. The \RNC quantity represents the charged-hadron multiplicity within the jet that includes the leading hadron.

In order to calculate \RT and \RNC, a high-\pT hadron is required, which is associated to the jet produced by a hard process. However, jettiness of an event can be defined in minimum-bias events as well, using transverse spherocity. It is defined as 
\begin{equation}
	\SO = \frac{\pi}{4} \min\limits_{\bf \hat{n}} \left( \frac{\sum_i \left| {\bf p}_{{\rm T},i} \times {\bf \hat{n}} \right| }{\sum_i p_{{\rm T},i}} \right)\ , 
\end{equation} 
where $i$ indexes the particles in the acceptance and ${\bf \hat{n}}$ is any unit vector in the azimuth plane. By construction, for jet-like events  $\SO \rightarrow 0$, while for isotropic events $\SO \rightarrow 1$.

Transverse spherocity concentrates on the central $\eta$ range and thus it is not sensitive to the part of event that expands toward higher $\eta$. To overcome this limitation, another event characterization variable called flattenicity ($\rho$) has been introduced recently~\cite{Ortiz:2022zqr}. This event quantifier is capable of selecting hedgehog-like events without a characteristic jetty structure in high-multiplicity pp collisions. The $\varphi$--$\eta$ plane is split up into roughly squarish cells of equal area, and the average transverse momenta of the charged particles is taken in each of them. Flattenicity is the relative standard deviation of the average momentum in a cell,
$$\rho = \frac{\sigma _{p_{\rm T}^{\rm cell}}}{\langle p_{\rm T}^{\rm cell} \rangle} .$$
Smaller $\rho$ corresponds to isotropic, and larger $\rho$ to more jetty events. Flattenicity has been succesfully used to classify events corresponding to the underlying physics process~\cite{Bencedi:2023iib,Ortiz:2022mfv,Horvath:2023lho}. By dividing the $\eta$ axis into 10 and the $\varphi$ axis into 8 ranges of equal length, we used 80 flattenicity cells altogether. 

The event-activity classes are summarized in Table~\ref{tab:class}. The limits were determined so that in most cases a similar number of events fell into each class.

\begin{table}[h]
	\begin{center}
		\begin{tabular}{ |c|c|c|c|c|c| } 
			\hline
			class & \#1 & \#2 & \#3 & \#4 & \#5 \\
			\hline\hline 
			\Nch & $\le$15 & 16--30 & 31--40 & 41--50 & $\ge$51 \\
			\Nfw & $\le$45 & 46--90 & 91--120 & 121--150 & $\ge$151 \\
			\RT & $<$0.5 & 0.5--1 & 1--1.5 & 1.5--2 & $>$2 \\
			\RNC & $<$0.5 & 0.5--1 & 1--1.5 & 1.5--2 & $>$2 \\
			\SO & 0--0.25 & 0.25--0.45 & 0.45--0.55 & 0.55--0.75 & 0.75--1 \\
			$\rho$ & 0--1 & 1--1.5 & 1.5--2 & 2--2.5 & $>$2.5 \\ 
			\Nmpi & $\le$5 & 6--10 & 11--13 & 14--16 & $\ge$17 \\ 
			\hline
		\end{tabular}
		\caption{\label{tab:class}Event classes for the event-activity observables \Nch, \Nfw, \RT, \RNC, \SO and $\rho$. For comparison, the \Nmpi categories are also shown.}
	\end{center}
\end{table}

\section{Charmed baryon-to-meson ratios}

\begin{figure}
	\includegraphics[width=\linewidth]{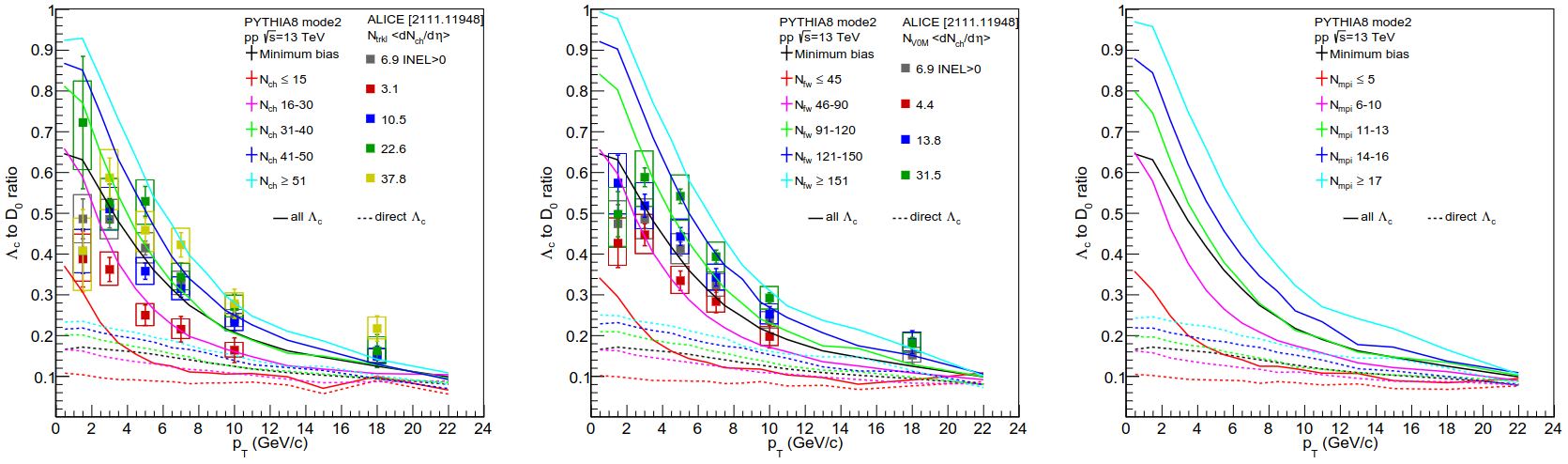}
	\caption{\LcToDz ratios from PYTHIA 8 soft-QCD simulations with the CR-BLC mode 2 in function of \pT, shown as solid lines, for charged-hadron multiplicity at mid-rapidity (\Nch, left), charged-hadron multiplicity at forward-rapidity (\Nfw, center), and number of multiparton-interactions (\Nmpi, right). The results for the \Nch and \Nfw classes are compared to data from ALICE~\cite{ALICE:2021npz}. The contribution of direct \Lc production is plotted separately as dashed lines.}
	\label{fig:NchNmpi}
\end{figure}

The left and center panels of Fig.~\ref{fig:NchNmpi} show the \LcToDz ratios in terms of \pT for different \Nch and \Nfw classes compared to data~\cite{ALICE:2021npz}, respectively. The right panel shows the \LcToDz ratios in terms of \pT in different \Nmpi categories. The simulations describe the trends both for central and forward pseudorapidity ranges, showing that the enhancement in the \LcToDz enhancement is stronger for higher event multiplicities, corresponding to a larger number of multiple-parton interactions.
The figure shows separately the direct \Lc production, excluding $\Sigc\rightarrow\Lc$. While the \Lc and \Sigc charm baryons have the same quark content, their isospins differ: for \Lc, I=0 and for \Sigc, I=1. One can note that the direct \Lc production has a much weaker \pT dependence than that from \Sigc decays, hinting that the excess charm-baryon production is linked to the isospin. A more detailed look at these two baryons is in the next section.

\begin{figure}
	\includegraphics[width=\linewidth]{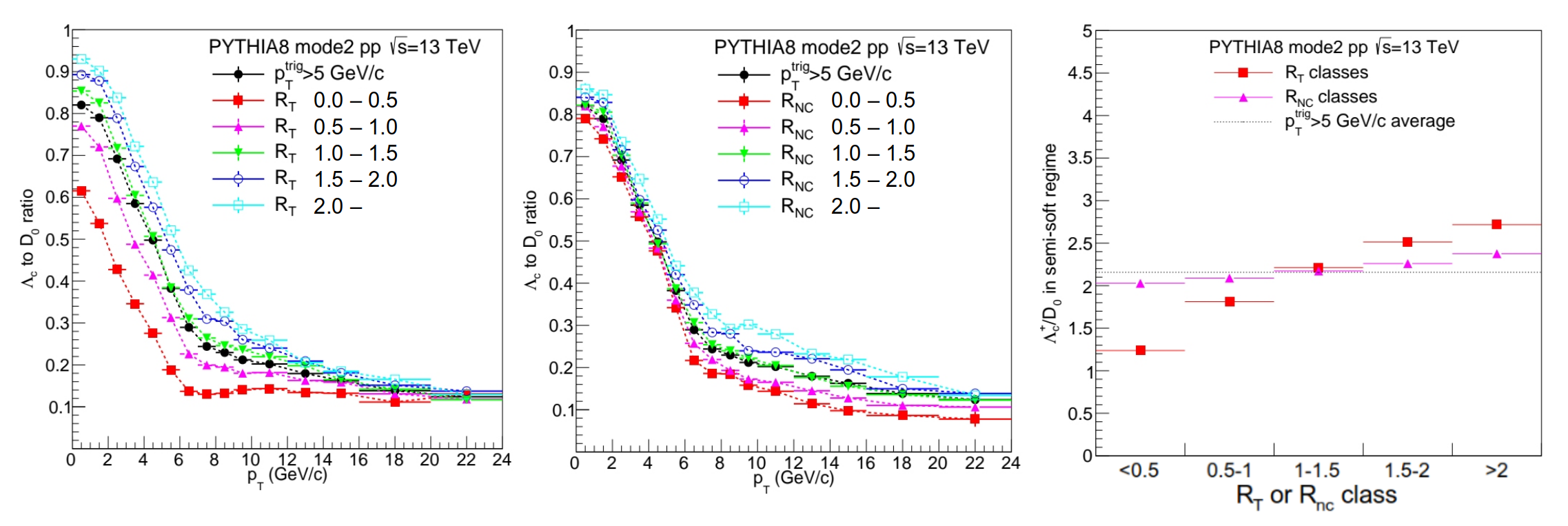}%
	\caption{\LcToDz ratios from PYTHIA 8 soft-QCD simulations with the CR-BLC mode 2 in events containing a $\pT>5$ GeV/$c$ hadron, in different \RT and \RNC classes. The left and center panels show \LcToDz in function of \pT in different \RT and \RNC classes, respectively. The right panel shows \LcToDz integrated over $2<\pT<8$ \GeVc as a function of \RT and \RNC bins, compared to the average (dashed line).}
	\label{fig:RtRnc}%
\end{figure}

The left and center panels of Fig.~\ref{fig:RtRnc} show the \LcToDz ratios in terms of \pT for different \RT and \RNC classes, respectively~\cite{Varga:2021jzb}. While higher \RT classes correspond to significantly higher enhancement in the low-to-mid transverse-momentum below $\pT \approx 10$ GeV/$c$, no such dependence can be observed on the \RNC classes in the same transverse-momentum range.
The right panel of Fig.~\ref{fig:RtRnc} shows the \LcToDz ratios  in function of the \RT and \RNC classes, integrated over the $2<\pT<8$ GeV/$c$, corresponding to the charm--light flavor coalescence range~\cite{Plumari:2017ntm},
The fact that the enhancement is a function of \RT but not of \RNC suggests that the excess production of $\Lc$ baryons is linked to the underlying event, and not to the jet within an event~\cite{Varga:2021jzb}. 

\begin{figure}
	\includegraphics[width=0.5\linewidth]{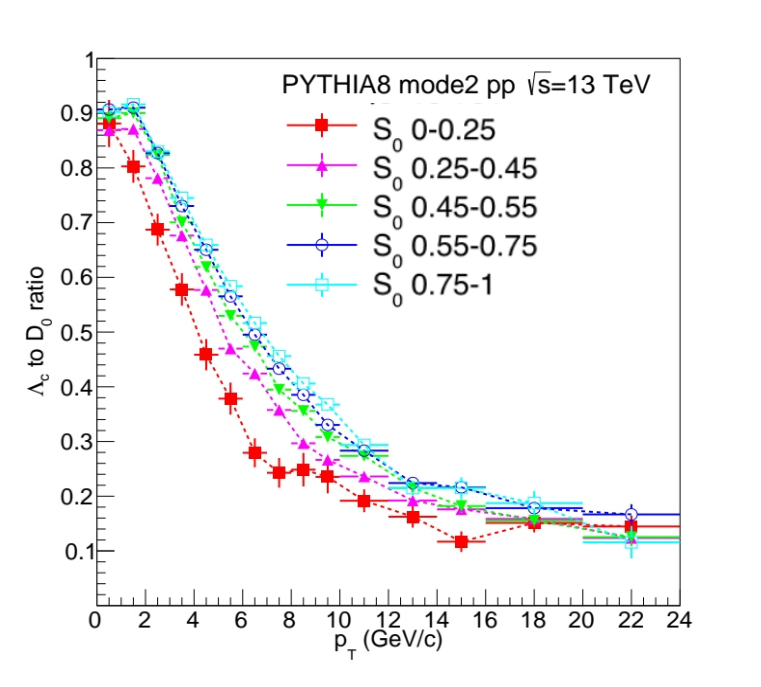}%
	\includegraphics[width=0.5\linewidth]{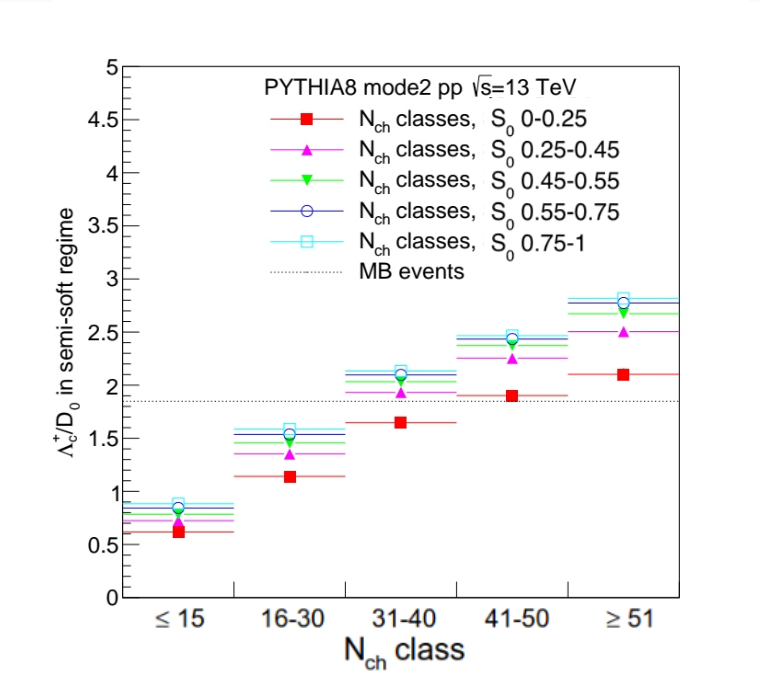}%
	\caption{\LcToDz ratios from PYTHIA 8 soft-QCD simulations with the CR-BLC mode 2, in different \SO classes. The left panel shows \LcToDz in function of \pT for $\Nch>50$. The right panel shows \LcToDz integrated over $2<\pT<8$ \GeVc as a function of \Nch, compared to the average (dashed line).}
	\label{fig:Sphero}%
\end{figure}

Fig.~\ref{fig:Sphero} (left) shows the \LcToDz ratio in function of the \pT for $\Nch > 50$ in different spherocity classes. A significant difference can be observed between the five spherocity classes. Fig.~\ref{fig:Sphero} (right) shows the \LcToDz values integrated for the coalescence regime $2<\pT<8$ GeV/$c$. While larger transverse spherocity corresponds to a stronger enhancement at fixed event multiplicity, the effect is more prominent toward high event multiplicities.

\begin{figure}
	\includegraphics[width=0.5\linewidth]{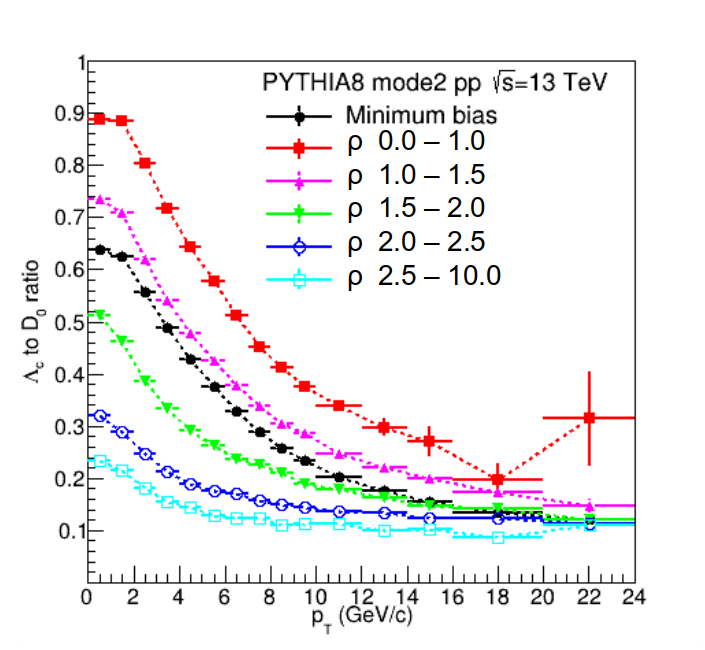}%
	\includegraphics[width=0.5\linewidth]{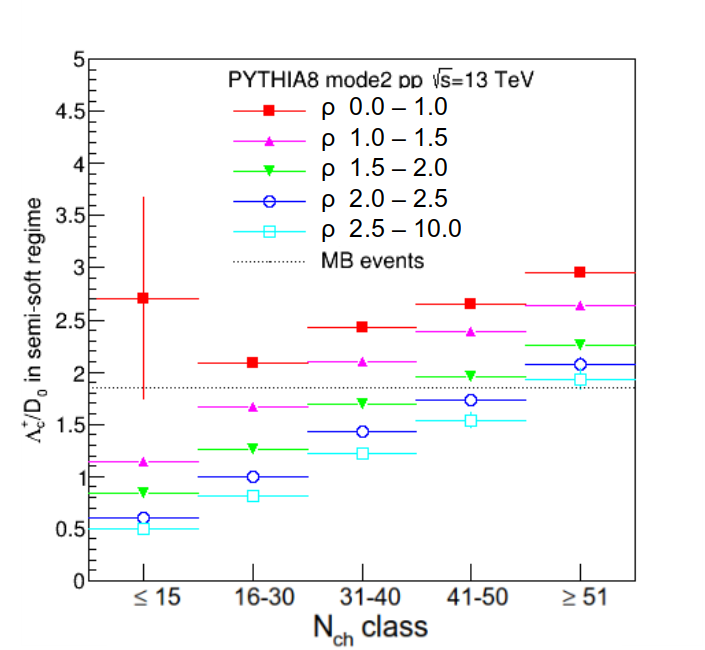}%
	\caption{\LcToDz ratios from PYTHIA 8 soft-QCD simulations with the CR-BLC mode 2, in different $\rho$ classes. The left panel shows \LcToDz in function of \pT. The right panel shows \LcToDz integrated over $2<\pT<8$ \GeVc as a function of \Nch, compared to the average (dashed line).}
	\label{fig:Flaten}%
\end{figure}

Fig.~\ref{fig:Flaten} (left) shows the \LcToDz ratio for different flattenicity classes, while Fig.~\ref{fig:Flaten} (right) shows the integrated enhancement for different flattenicity classes for several \Nch classes. Flattenicity correlates with \Nmpi and therefore it is strongly bound to the underlying event. The enhancement in \LcToDz ratio decreases with increasing flattenicity, in every \Nch class, which makes a flattenicity more sensitive than transverse spherocity. It is also more powerful than \RT and \RNC in the sense that events not containing a high-\pT hadron can also be analyzed. Future measurements of the charmed baryon and meson production in terms of flattenicity will therefore provide crucial feedback for models and will therefore play key role in the understanding of heavy-flavor fragmentation.

\section{Charmed baryon-to-baryon ratios}

\begin{figure}
	\includegraphics[width=\linewidth]{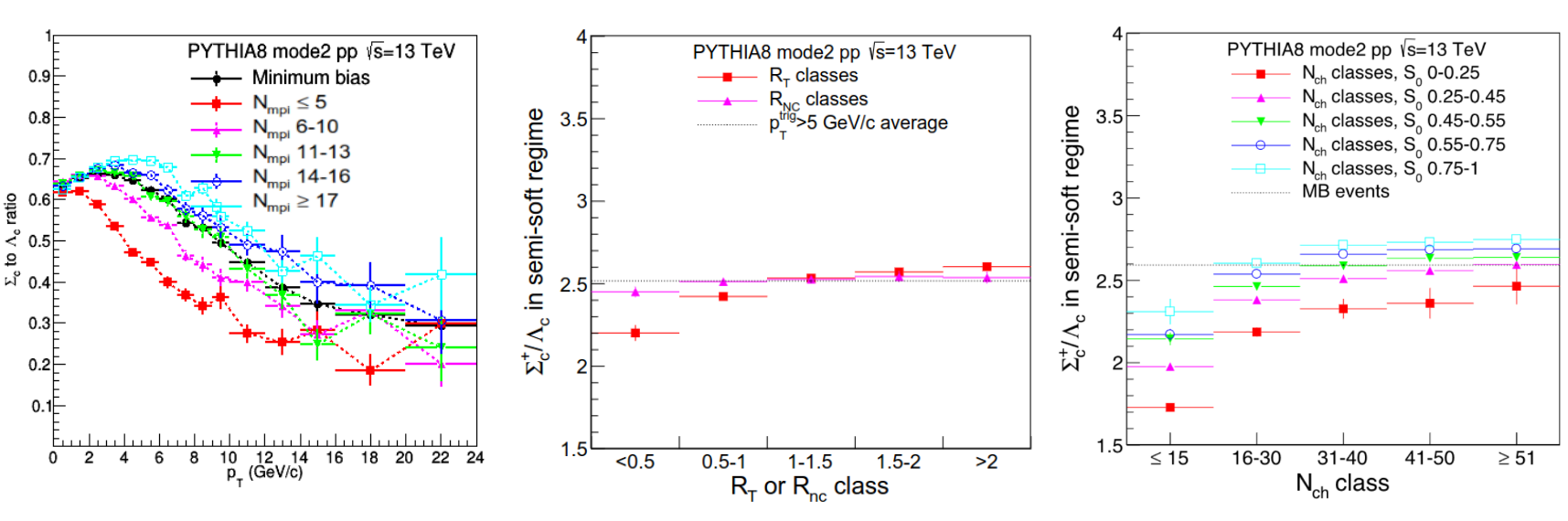}%
	\caption{$\Sigc/\Lc$ ratios from PYTHIA 8 soft-QCD simulations with the CR-BLC mode 2. In the left panel, $\Sigc/\Lc$ is shown for different \Nmpi ranges in function of \pT. In the center panel $\Sigc/\Lc$ is shown integrated over $2<\pT<8$ \GeVc as a function of \RT and \RNC bins, compared to the average (dashed line). The right panel shows $\Sigc/\Lc$ integrated over $2<\pT<8$ \GeVc in different \SO classes, as a function of \Nch, compared to the average (dashed line).}
	\label{fig:SigcLc}%
\end{figure}

The ratios of different charmed baryons are sensitive observables that depend on hadronization mechanisms~\cite{ALICE:2022cop}. The ratio of charmed baryons $\Sigc/\Lc$ is shown in Fig.~\ref{fig:SigcLc} (left) for different \Nmpi categories. Although these two baryons have the same quark content and only differ in their isospin, a substantial difference is observed between the enhancement trends of \Lc and \Sigc, which depends on the number of multiparton-interactions. From the \Nmpi dependence one also expects a difference with respect to the \RT and \RNC, as well as \SO classes. The $\Sigc/\Lc$ enhancement for different \RT and \RNC classes is shown for the coalescence regime $2<\pT<8$ GeV/$c$ in Fig.~\ref{fig:SigcLc} (center), and for different transverse spherocity and multiplicity classes (right). The experimental verification of this dependence on the event shapes will shed light on the fragmentation mechanism of heavy quarks. 

\begin{figure}
	\includegraphics[width=\linewidth]{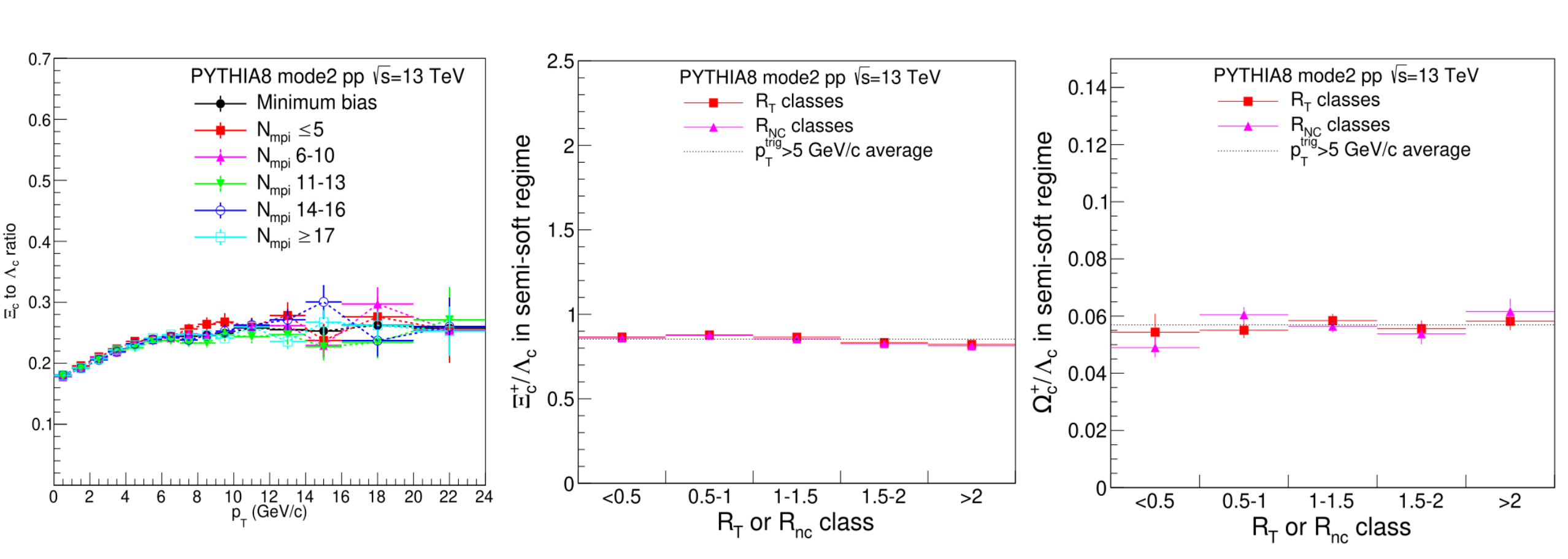}%
	\caption{$\Xic/\Lc$ and $\Omc/\Lc$ ratios from PYTHIA 8 soft-QCD simulations with the CR-BLC mode 2. In the left panel, $\Xic/\Lc$ is shown for different \Nmpi ranges in function of \pT. In the center and right panels, $\Xic/\Lc$ and $\Omc/\Lc$ are shown, respectively, integrated over $2<\pT<8$ \GeVc as a function of \RT and \RNC bins, compared to the average (dashed line).}
	\label{fig:XicOmc}%
\end{figure}

Fig.~\ref{fig:XicOmc} (left) shows the $\Xic/\Lc$ ratio for different \Nmpi categories. 
The central and right panels of Fig.~\ref{fig:XicOmc} show the integrated  $\Xic/\Lc$ and $\Omc/\Lc$ ratios, respectively, for different \RT and \RNC classes and integrated in the  coalescence regime $2<\pT<8$ GeV/$c$. 
The increasing trend with the transverse momentum shows that while the low-\pT enhancement can be linked to charm, there is a stronger enhancement from strangeness content in the high-\pT regime. A similar trend has been observed for the $\Omc/\Lc$ ratio~\cite{Varga:2023byp}. 
However, the dependence of the enhancement is not linked to the event activity, as the curves corresponding to different \Nmpi classes are on top of each other in the left panel, and there is no \RT or \RNC dependence present in the two right-hand-side panels. Hence it is likely that the charm baryon enhancement driven by a different mechanism than strange baryon enhancement, at least in the frames of the model class used in this analysis.

\section{Conclusions}

An enhancement of \LcToDz has been observed in pp collisions at the LHC, compared to that computed from e$^+$e$^-$ collisions using the fragmentation hypothesis. This questions the universality of charm fragmentation.
We proposed event-activity classifiers which provide great sensitivity to the production mechanisms. These provide directly accesible experimental observables in the ongoing LHC Run 3 data collection phase. In a model class with color reconnection beyond leading approximation, the \Lc enhancement comes from the underlying event, not from the jets. This is seen by comparing the behavior of the enhancement in classes of transverse and jet event-activity in high-\pT hadron triggered events as well as in transverse spherocity classes in minimum bias events.
We found that the usage of flattenicity, a new quantity to represent multiple-parton interactions, may be even more distinctive and help pin down which types of events the enhancement stems from.
The above-mentioned observables are sensitive to differences between mechanisms of strangeness and charm enhancement as well as baryon isospin.
Comparing our results to future experimental measurements on the event-activity-dependent properties of charmed-baryon production will provide invaluable feedback for the understanding of charm production and hadronization mechanisms in pp collisions. These, in turn, will provide the baseline for the exploration of heavy-ion collisions.

\section*{Acknowledgements}

This work has been supported by the NKFIH grants OTKA FK131979 and K135515, as well as by the 2021-4.1.2-NEMZ\_KI-2022-00007 project. The author acknowledges the computational resources provided by the Wigner GPU Laboratory and the research infrastructure provided by the Hungarian Research Network (HUN-REN). The authors are grateful for the support of NPP Paks.


\bibliographystyle{utphys}
\bibliography{LcDbiblio}

\end{document}